\documentclass{mn2e}
\usepackage{psfig,rotating}

\newcommand{\chandra}{{\it Chandra}}
\newcommand{\asca}{{\it ASCA}}
\newcommand{\rosat}{{\it ROSAT}}
\newcommand{\sax}{{\it BeppoSAX}}
\newcommand{\xmm}{{\it XMM-Newton}}
\newcommand{\einstein}{{\it Einstein}}
\newcommand{\lum}{\thinspace\hbox{$\hbox{erg}\thinspace\hbox{s}^{-1}$}}

\newcommand{\ic}{IC\,342}

\title[{\it X-ray population of IC\,342}]
{{\it XMM-Newton} observation of the X-ray point source
population of the starburst galaxy IC\,342}
\author[A. K. H. Kong]{Albert K. H. Kong\thanks{E-mail:
akong@cfa.harvard.edu}\\
Harvard-Smithsonian Center for Astrophysics, 60 Garden Street, Cambridge,
MA 02138, USA}
\begin{document}

\date{Accepted. Received; in original form 2003 February 20}

\pagerange{\pageref{firstpage}--\pageref{lastpage}} \pubyear{2003}

\maketitle

\label{firstpage}

\begin{abstract}
We present the results of an \xmm\ observation of the starburst galaxy
\ic. Thirty-seven X-ray point sources were detected down to a luminosity
limit of $\sim 10^{37}$ \lum. Most of the sources are located near the
spiral arms. The X-ray point source luminosity function is consistent
with a power-law shape with a slope of $0.55$, typical of starburst
galaxies.   
We also present the energy spectra of several ultraluminous X-ray sources
(ULXs), including the luminous X-ray source in the galactic nucleus. Except
for the nucleus and a luminous supersoft X-ray source, other ULXs can
generally be fit with a simple power-law
spectral model. The nucleus is very luminous ($\sim 10^{40}$\lum\ in
0.2--12 keV) and 
requires disc blackbody and power-law
components to describe the X-ray emission. The spectral fit reveals a
cool accretion disc ($kT=0.11$ keV) and suggests that the source
harbours either an intermediate-mass black hole or a stellar-mass black
hole with outflow.

\end{abstract}

\begin{keywords}
galaxies: individual: \ic\ -- X-rays: binaries -- X-rays: galaxies.
\end{keywords}

\section{Introduction}

\ic\ is a nearby late-type Sc galaxy in the Maffei Group which is one
of the closest groups to our Galaxy. The spiral arms of \ic\ is well developed
and is almost face-on ($i=25^{\circ}\pm3^{\circ}$; Newton 1980). \ic\
also shows substantial nuclear star formation (Becklin et al. 1980;
Rickard \& Harvey 1984). \ic\ has been observed with \einstein\
(Fabbiano \& Trinchieri 1987), \rosat\ (Bregman, Cox, \& Tomisakam 1993)
and \asca\ (Okada et al. 1998; Kubota et al. 2001)
previously. Ten X-ray point sources were found in the \rosat\
HRI observation. 
\ic\ is also well known to house several ultraluminous X-ray sources
(ULXs; Roberts \& Warwick 2000). In particular, X1 and X2 (based on
designations in Okada et al. 1998) showed spectral
and intensity transition during the \asca\ observations
taken in 1993 and 2000 (Kubota et al. 2001). The state transitions of
the two ULXs resemble the spectral/intensity states of X-ray
transients in our Galaxy.
In addition, a periodicity of 31 hr or 41 hr was found in X-2 with long
($\sim 250$ks net exposure time) \asca\ observations (Sugiho et
al. 2001). Although we have known that there are several point sources
around the galaxy nucleus based on \rosat\ HRI data, the angular
resolution of \asca\ is very low and the nucleus of \ic\ is not
resolved. Hence, we know very little about the X-ray emission from the
nucleus. With the advent of \xmm, we now have sufficient resolving
power and collective area to have a detailed study of the galaxy.

At a distance of 1.8 Mpc (see Buta and McCall 1999 for a review), $1''$
corresponds to 8.7 pc. Its proximity and
its almost face-on orientation towards the observer provides a unique
possibility to study the X-ray point source populations. Unfortunately,
\ic\ is located at low Galactic latitude ($b=10^{\circ}.6$), resulting a
relatively high hydrogen column density, $N_H=3\times10^{21}$
cm$^{-2}$ (Dickey \& Lockman 1990). This limits us to constrain local
absorption and X-ray emission below 1 keV.

In this paper, we report on the X-ray point sources detected in \ic,
with an archival \xmm\ observation. We begin with a description of the
observation and data reduction procedures. In Section 3, we present
the results of our study including the global properties of X-ray
sources, spectral fits to bright sources, and the X-ray luminosity
function. In Section 4, we discuss the X-ray source population, 
the nucleus and the ULXs of this galaxy. A summary will be given in
Section 5.


\section{Observations and data reduction}

\ic\ was observed with the three instruments of European Photon Imaging Camera (EPIC) and
the Optical Monitor (OM) on board \xmm\ on 2001 February 11 for about
10ks. The instrument modes were full-frame with medium optical
blocking filter for the pn
and the two MOS cameras, while the filter of OM was $UVW1$ (bandpass
between 2200$\mathring{\rm A}$ and 4000$\mathring{\rm A}$). The event
files were reprocessed and filtered with the \xmm\ Science Analysis
Software (SAS v5.4.1).  Only data in 0.2-12 keV were used for analysis. 
We constructed light curves of source-free regions from each of the
three instruments and 
after rejecting intervals with a high background level, we obtained a good
time interval of 5ks and 9.5ks for the pn and MOS cameras, respectively.

We created a combined MOS image by merging the two MOS
detectors (MOS-1 and MOS-2) to increase the signal-to-noise (S/N)
level. We also obtained a merged image from the three
detectors. Figure 1 shows the merged EPIC image of the central
$20'\times20'$ region of \ic. Source
detection was done on the pn, merged MOS, and pn+MOS images with SAS task 
{\sc EWAVELET}, with a detection threshold of $5\sigma$. We examined the
existence and position of each detected source by checking individual and
merged images. The OM image and source detection were performed with the
standard SAS pipeline tasks. We also corrected the astrometry of the OM
image by using the USNO catalogue (Monet et al. 1998). The OM image of the
central $8'\times8'$ region of \ic\ is shown in Figure 2.

Source count rates were determined via aperture photometry with
source-free regions as the background and were corrected for
effective exposure and vignetting. The
radius of the aperture was varied with average off-axis angle in order
to match the 90\% encircled energy function. 
The count rates reported here are
the pn count rates in the total (0.2-12 keV), soft (0.2-1 keV), medium
(1-2 keV), and high (2-12 keV) bands. Some sources fell in the chip
boundary or outside the field of view of the pn camera. We then used the
MOS count rates and rescaled them to the pn values with PIMMS.  
Energy spectra of bright sources were extracted with the SAS task
{\sc XMMSELECT}. Source-free
regions were used for background. Response matrices were created by
{\sc RMFGEN} and {\sc ARFGEN}. 

\section{Results}

\subsection{Properties of X-ray sources}

We detected 37 X-ray point sources in \ic\ (Table 1). 
The sources are mainly concentrated in the
spiral arms of the galaxy (see Fig. 3). Five of the detected sources were
clearly associated with bright stellar object within $5''$ in the
Digitized Sky Survey (DSS; Fig. 3) and the OM (Fig. 2) images. Source
12 (X12 hereafter) was 
also previously associated with a star in the \rosat\ observation (Source 
4 in Bregman et al. 1993). We then
used the 3 matches within the central $5'$ region to correct the astrometry
of the X-ray images with the USNO catalogue. The
positions listed in Table 1 and throughout this paper use this astrometric
reference. The conversion of luminosities assumes an absorbed power-law
spectrum with a photon index of 2 and $N_H=8\times10^{21}$ cm$^{-2}$
(the average from the spectral fits; see \S\,3.2). The detection
limit is $\sim 1.3\times10^{37}$ \lum\ in the 0.2--12 keV band. Of the 37
sources, 9 of them were detected in previous \rosat\ HRI observation
(Bregman et al. 1993). One \rosat\ source (source 2 in Bregman et
al. 1993) is below the \xmm\ detection limit, which suggests that the
source is fainter by a factor of $\sim 7$ between the two observations.

\begin{figure}
\psfig{file=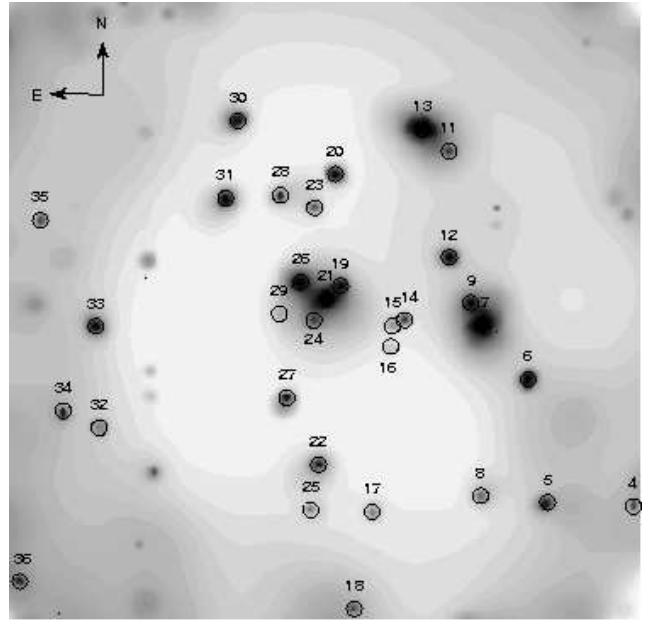,width=8.4cm}
\caption{Merged EPIC (pn + MOS-1 + MOS-2) image (0.2-12 keV) of the central
$20'\times20'$ region of \ic. The image has been exposure corrected
and adaptively smoothed. The circles have a radius of $15''$. Some
apparent ``sources'' (which are not included in the source list) in
the image are artifacts of the detector.}
\end{figure}

\begin{figure}
\psfig{file=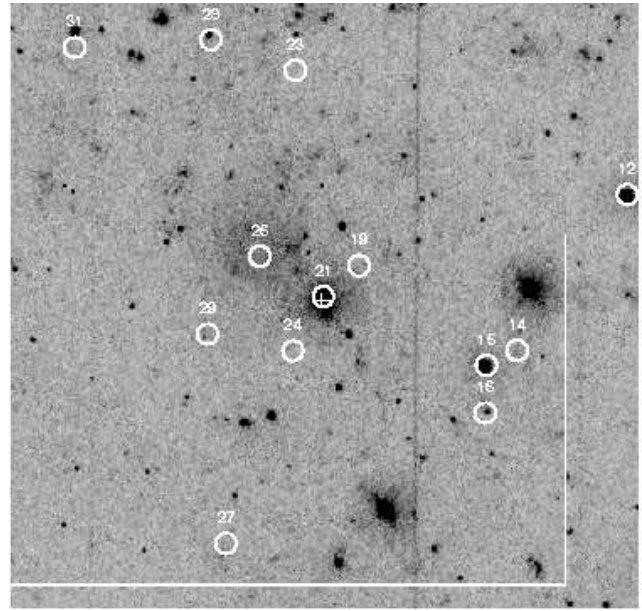,width=8.4cm}
\caption{Optical Monitor (OM) image of the central $8'\times8'$ region
of \ic\ in the $UVW1$ filter. The
circles (radius of $8''$) correspond to the X-ray sources identified
from the \xmm\ images.
The dynamical centre of the galaxy (Turner et al. 1992) is marked with a
cross which is coincident with X21.}
\end{figure} 

\begin{figure*}
\psfig{file=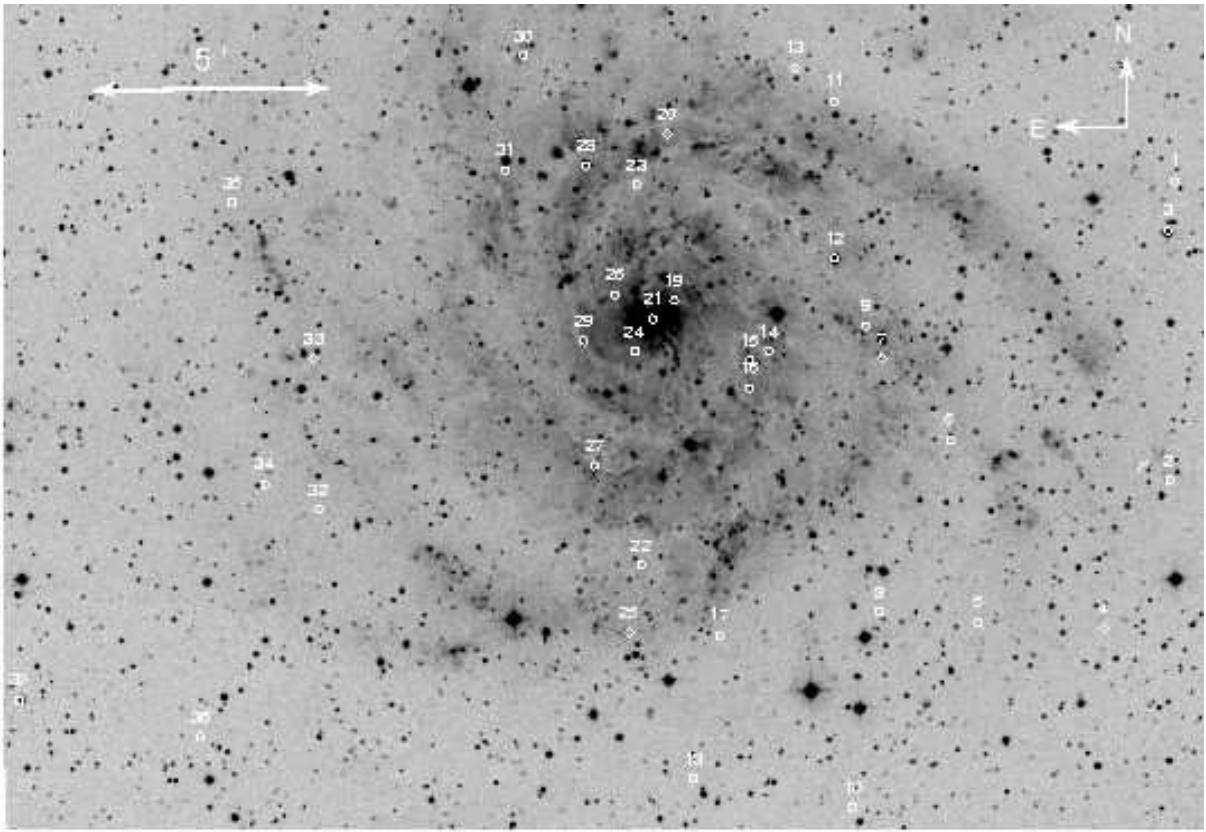,width=16cm}
\caption{Digitized Sky Survey (DSS) blue band image of \ic, with
detected \xmm\ X-ray sources overplotted (see Table 1 for the source
list). The circles have a radius of $5''$.}
\end{figure*} 

Many of the detected sources have $< 100$ counts, which makes it difficult
to derive spectral parameters. However, hardness ratios can give a crude  
indication of the X-ray spectra. Following Kong et al. (2002), we computed
the hardness ratios for each detected sources, which are defined as      
HR1=(medium-soft)/(medium+soft) and HR2=(hard-soft)/(hard+soft). Figure 4
shows the colour-colour diagram for all detected sources. We have overlaid
the colour-colour diagram with lines showing the tracks followed by   
representative spectral models with differing values of $N_H$. We have
shown in the diagram for power-law models with photon index ranges from
1.2 to 3, a Raymond-Smith thermal model with $kT_{RS}=0.5$ keV, and a
blackbody model with $kT=0.1$ keV. For each model, the $N_H$
increases from the left to the right, from $N_H=3\times10^{21}$
cm$^{-2}$ to $N_H=10^{22}$ cm$^{-2}$. Power-law spectra
tend to occupy the top right section of the diagram, while soft thermal
models occupy the lower left.

\begin{figure}
\psfig{file=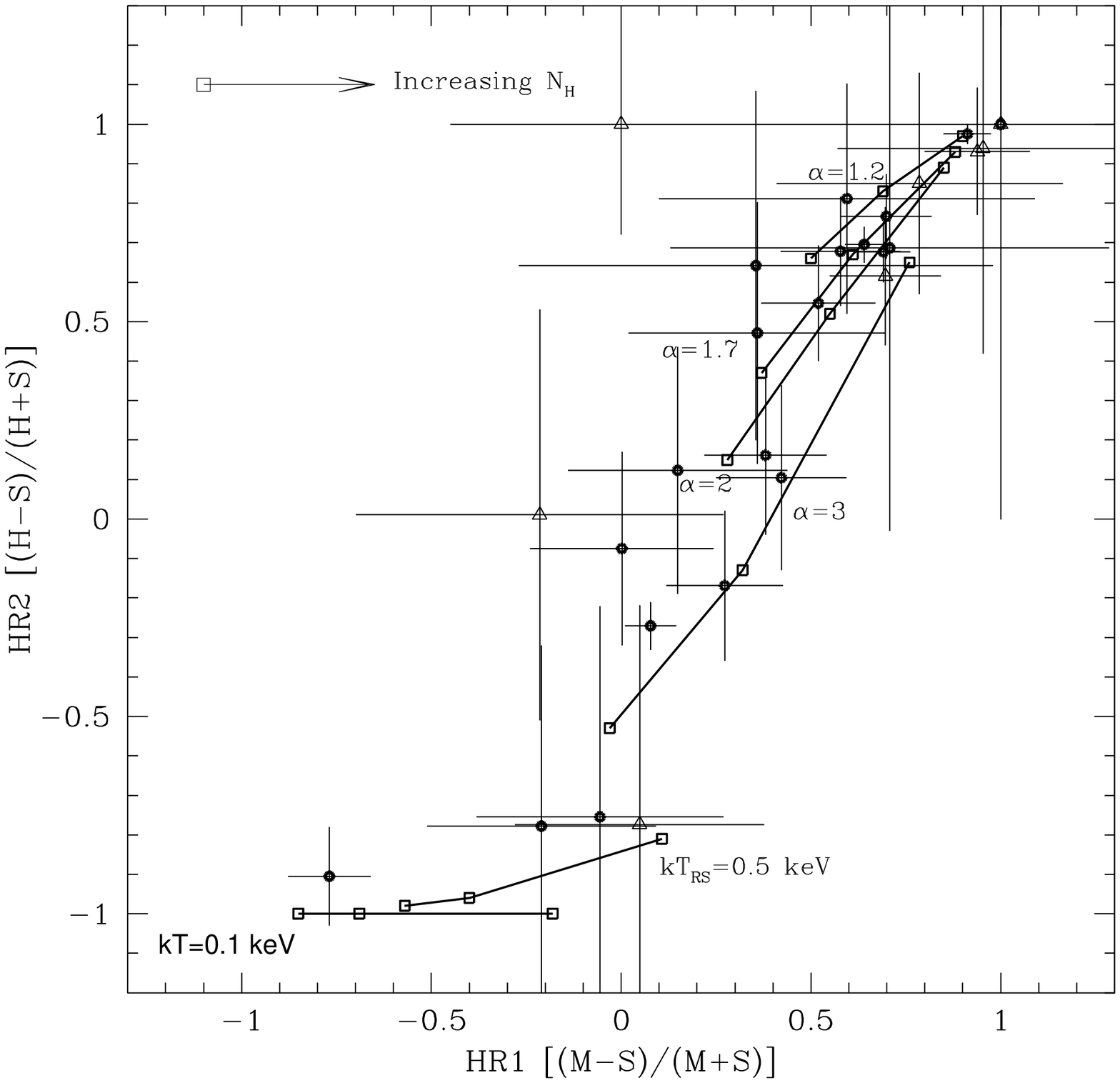,width=8.6cm}
\caption{Colour-colour diagram for all detected sources excluding the
foreground stars (circles: sources
within the $D_{25}$ ellipse; triangles: sources outside the $D_{25}$
ellipse). Also plotted 
are the hardness ratios estimated from different spectral
models. Top to bottom: Power-law model with $\alpha$ of 1.2, 1.7, 2,
and 3, Raymond-Smith model with $kT_{RS}$ of 0.5 keV, and blackbody
model with $kT$ of 0.1 keV. For each model, $N_H$ varies from the left
from $3\times10^{21}$cm$^{-2}$, $5\times10^{21}$cm$^{-2}$, and $10^{22}$cm$^{-2}$.}
\end{figure}

It is clear that many of the sources are consistent with a power-law 
spectral model, while a few of them are dominated by soft X-ray emission.
These soft X-ray sources are likely to be supersoft sources (Di\,Stefano
\& Kong 2003), supernova remnants or foreground stars. For example,
foreground star X12 has HR1=$-0.58$ and HR2=$-0.90$.

The closest X-ray source to the dynamical galaxy centre (Turner et
al. 1992; marked with a
white cross in Figure 2), X21, is about $3''$ (26.1 pc) from the galactic
centre and is within the positional accuracy of \xmm. 
It it worth noting that X21 is also seen in the
\rosat\ HRI image, but in the \einstein\ and \asca\ images, X21 is not
resolved from several nearby sources (see Figure 1).
The galaxy centre was also detected as a point-like UV source in the
OM, with $UVW1=13.6$.

\begin{table*}
\caption{\xmm\ source list of \ic}
\small
\begin{tabular}{lccccccl}
\hline
ID & R.A. & Dec. & \multicolumn{3}{c}{Net Counts} & $L_X$$^a$ & Note\\
   & (J2000.0) & (J2000.0)  & 0.2--1 keV & 1--2 keV & 2--12 keV
&($\times10^{38}$) &\\
\hline
X1& 03:44:47.5 & +68:08:45.6 & $52.0\pm10.1$&$290.4\pm18.9$&$218.6\pm18.0$&1.11&\\
X2& 03:44:49.4 & +68:02:17.5 & $0.0\pm4.9$&$108.7\pm12.0$&$167.4\pm16.0$&0.53&\\
X3& 03:44:49.3 & +68:07:42.1 & $38.9\pm10.8$&$59.4\pm9.9$
&$-2.7\pm4.4$ &0.25 & $B=13.1$, $R=12.3$\\
X4& 03:45:04.6 & +67:59:05.6 & $-2.682\pm3.3$&$11.1\pm4.6$&$9.1\pm6.6$&0.14&\\
X5& 03:45:33.6 & +67:59:14.4 & $6.7\pm6.4$&$222.6\pm16.8$&$1.7\pm6.2$&0.84&\\
X6& 03:45:39.9 & +68:03:11.6 &
$58.1\pm23.6$&$327.7\pm18.9$&$113.8\pm11.8$&1.64&\rosat\#1 \\
X7& 03:45:55.4 & +68:04:57.6 &
$194.0\pm14.4$&$885.0\pm29.9$&$1079.6\pm33.2$&16.74&\rosat\#3 \\
X8& 03:45:56.4 & +67:59:28.2 & $3.8\pm3.8$&$8.0\pm3.9$&$17.6\pm6.1$&0.23&\\
X9& 03:45:59.2 & +68:05:39.9 & $18.5\pm5.1$&$69.3\pm8.6$&$96.8\pm10.4$&1.43&\\
X10& 03:46:02.6 & +67:55:14.2 & $11.2\pm5.7$&$7.2\pm3.8$&$11.4\pm6.5$&0.21&\\
X11& 03:46:06.3 & +68:10:32.3 & $3.5\pm4.1$&$13.8\pm4.6$&$33.9\pm7.5$&0.40&\\
X12& 03:46:06.5 & +68:07:07.8 & $117.2\pm11.4$&$31.5\pm6.1$ &
$5.9\pm4.7$&1.20 & $B=12.4$, $R=12.5$, \rosat\#4 \\
X13& 03:46:15.4 & +68:11:15.6 &
$14.7\pm5.0$&$320.3\pm18.0$&$1178.4\pm34.6$&11.79& \rosat\#5 \\
X14& 03:46:21.8 & +68:05:07.7 & $0.0\pm1.3$&$115.5\pm11.1$&$3.1\pm2.8$&0.34&\\
X15& 03:46:26.0 & +68:04:56.2 & $16.9\pm4.5$&$16.8\pm4.3$ &$1.9\pm2.6$
&0.26 & $B=13.5$, $R=14.1$\\ 
X16& 03:46:26.2 & +68:04:19.4 & $10.6\pm4.0$&$13.9\pm4.1$ &$4.4\pm3.9$
&0.23 & $B=16.7$, $R=16.6$\\
X17& 03:46:33.1 & +67:58:57.9 & $1.9\pm3.8$&$11.6\pm4.3$&$10.7\pm6.0$&0.19&\\
X18& 03:46:39.2 & +67:55:51.9 & $-5.6\pm2.0$&$-6.4\pm1.2$&$31.3\pm8.0$&0.15&\\
X19& 03:46:43.5 & +68:06:13.6 &
$22.4\pm5.1$&$71.0\pm8.5$&$76.6\pm9.1$&1.32&\rosat\#6 \\
X20& 03:46:45.1 & +68:09:49.7 &
$98.1\pm10.3$&$12.8\pm4.4$&$4.8\pm5.1$&0.90&\rosat\#7 \\
X21& 03:46:48.4 & +68:05:49.6 & $316.3\pm17.9$&$369.5\pm20.0$
&$181.5\pm13.8$ &6.74 & nucleus, \rosat\#8 \\
X22& 03:46:51.2 & +68:00:30.0 & $-1.3\pm3.4$&$27.2\pm5.8$&$29.0\pm6.8$&0.43&\\
X23& 03:46:52.2 & +68:08:44.8 & $-2.7\pm2.9$&$17.2\pm5.0$&$16.2\pm5.4$&0.24&\\
X24& 03:46:52.7 & +68:05:07.7 & $22.6\pm5.3$&$22.7\pm5.0$&$19.5\pm5.3$&0.51&\\ 
X25& 03:46:53.8 & +67:59:02.3 & $18.5\pm5.6$&$12.1\pm4.0$&$2.3\pm4.7$&0.26&\\
X26& 03:46:57.2 & +68:06:21.1 &
$197.7\pm16.8$&$1083.3\pm33.1$&$204.2\pm14.7$&4.56&\rosat\#9 \\
X27& 03:47:01.9 & +68:02:38.6 & $16.4\pm5.4$&$22.2\pm5.4$&$21.1\pm6.2$&0.45&\\
X28& 03:47:03.9 & +68:09:08.7 & $9.0\pm4.7$&$19.2\pm4.9$&$25.2\pm6.6$&0.42&\\
X29& 03:47:04.4 & +68:05:20.8 & $14.2\pm4.9$&$12.7\pm4.1$&$1.9\pm4.4$& 0.23&\\
X30& 03:47:18.4 & +68:11:31.9 & $41.2\pm7.3$&$72.1\pm9.0$&$29.3\pm7.1$&1.12&\\
X31& 03:47:22.7 & +68:09:02.6 & $28.1\pm6.5$&$69.2\pm8.9$&$34.6\pm7.4$&1.02&\\
X32& 03:48:05.6 & +68:01:41.0 & $45.4\pm8.7$&$50.1\pm9.7$&$-6.2\pm4.3$&0.20&\\
X33& 03:48:06.8 & +68:04:56.1 &
$31.3\pm6.5$&$69.8\pm8.7$&$43.4\pm7.9$&1.13&\rosat\#10 \\
X34& 03:48:17.9 & +68:02:13.2 & $0.5\pm3.7$&$22.6\pm5.2$&$17.2\pm6.6$&0.30&\\
X35& 03:48:25.9 & +68:08:19.6 & $-0.4\pm3.9$&$9.0\pm4.2$&$15.5\pm6.0$&0.19&\\
X36& 03:48:32.3 & +67:56:43.9 & $8.4\pm6.4$&$70.3\pm11.9$&$10.9\pm7.0$&0.36\\
X37& 03:49:14.0 & +67:57:30.0 & $8.1\pm4.2$&$42.8\pm7.1$ &$43.0\pm8.3$
& 0.74& $B=14.0$, $R=13.0$\\
\hline
\end{tabular}

X-ray sources associated with stars are indicated by the $B$ and $R$
magnitudes taken from the USNO catalogue. Sources
detected by\\
\hspace*{-11.3cm}\rosat\ HRI (Bregman et al. 1993) are noted.\\
$^a$ Unabsorbed 0.2--12 keV luminosity in unit of \lum,
by assuming a power-law spectrum with $N_H=8\times10^{21}$ cm$^{-2}$
and $\alpha=2$. 
\end{table*}

\subsection{Spectral fits of bright sources}

We extracted the energy spectra for sources that have $> 100$ counts, and
fitted them to simple one-component spectral models including absorbed
power-law, blackbody, disc blackbody, Raymond-Smith, broken powerlaw and the bulk
motion Comptonization (BMC; Shrader \& Titarchuk 1999) models.  Except for X12, X20, and
X21, all sources were satisfactorily fitted with simple absorbed power-law
model (see Table 2). Generally, the $N_H$ of the
brightest sources ranges
from $3\times10^{21}$cm$^{-2}$ to $2\times10^{22}$cm$^{-2}$ with an
average of $8\times10^{21}$ cm$^{-2}$, while
the photon index varies between 1.7 and 3.2. These spectra are consistent with typical X-ray binaries seen in
our Galaxy and other external galaxies. For X7 and X13, the spectra
can be fitted equally well with a disc blackbody model with inner
disc temperature of $\sim 2$ keV. Figure 5 shows the disc blackbody
spectrum of X7.
Both X7 and X13 are two of the previously reported
ULXs in \ic\ and our results are in good agreement with previous
\asca\ observations (see \S\,4.3 for discussion).
For X12 and X20, their spectra are soft and can be
fit with a blackbody model. X12 is likely to be a foreground star and
the soft spectrum is not unexpected.
The blackbody temperature of X20 is $\sim 0.1$ keV, suggesting that it is a luminous supersoft
X-ray sources (Di\,Stefano \& Kong 2003). It is worth noting that X20
was also detected by \rosat\ HRI (Bregman et al. 1993). If we assumed
the same spectral shape as the \xmm\ observation, X20 is about a
factor 2 brighter during the \rosat\ observations.

X21 is about $3''$ from the dynamical centre and probably is the nucleus of
the galaxy (see \S\,4.2 for discussion). Single-component spectral models cannot give a good fit
(see Table 2)
and it requires a more complicated model.
We fit X21 with a two-component model
consisting of disc blackbody and power-law components. The additional
component significantly ($>$ 99.9\% confidence) improves the fit. We
also considered the BMC model and the spectral fit is comparable to
the disc blackbody + power-law model. The blackbody temperatures of
both models are similar and are low ($kT\approx0.1$ keV) compared with
many ULXs. Depending on the spectral model, the 0.2--12 keV luminosity is between
$6\times10^{39}$\lum\ and $1.6\times10^{40}$\lum.
The best-fitting disc blackbody + power-law model is shown in Figure 6. 

\begin{figure}
\rotatebox{-90}{\psfig{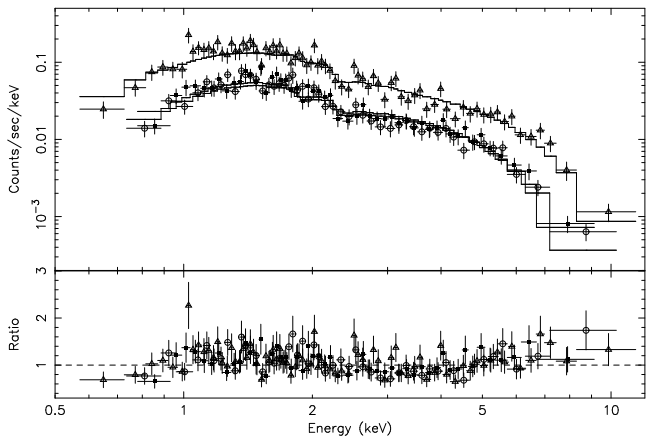}}
\caption{Energy spectrum of X7 (\ic\ X-1 in Okada et
al. 1998). pn, MOS-1, and MOS-2 data are marked as triangles, solid
squares and circles, respectively. The spectrum can be fitted with a
disc blackbody model with $N_H=3.2\times10^{21}$ cm$^{-2}$ and
$kT_{in}=1.98$ keV.}
\end{figure}

\begin{figure}
\rotatebox{-90}{\psfig{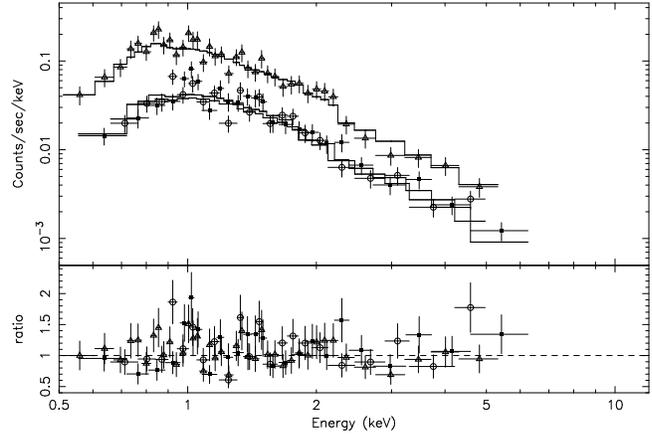}}
\caption{Energy spectrum of the nuclear source (X21). pn, MOS-1, and
MOS-2 data are marked as triangles, solid 
squares and circles, respectively. The spectrum can be fitted with a
disc blacbody + power-law model with $N_H=8.7\times10^{21}$ cm$^{-2}$, $\alpha=2.73$, and
$kT_{in}=0.11$ keV.}
\end{figure}

\begin{table}
\caption{Spectral fits to the brightest X-ray sources in \ic}

\footnotesize
\begin{tabular}{lccccc}
\hline
ID & $N_H$& $\alpha$ & kT & $\chi^2_{\nu}$/dof & $L_X$$^a$\\
       &($10^{22}$cm$^{-2}$)& & (keV) & & \\
\hline

X6 &$1.32^{+0.90}_{-0.71}$ & $2.53^{+1.03}_{-0.73}$ & & 0.4/7 &$2.23$\\
X7 &$0.60^{+0.06}_{-0.05}$ & $1.72^{+0.08}_{-0.08}$ & & 0.81/174 & $13.6$\\
   &$0.32^{+0.03}_{-0.04}$ & & $1.98^{+0.15}_{-0.13}$$^b$& 1.03/174 &$10.4$\\
X9 &$0.93^{+0.91}_{-0.72}$ & $1.84^{+0.56}_{-0.38}$ & & 0.5/3 & $1.45$\\
X12&$0.16^{+0.28}_{-0.13}$ & &$0.17^{+0.06}_{-0.05}$ & 1.48/10 &$0.29$\\
X13 & $2.94^{+0.48}_{-0.43}$ & $2.16^{+0.23}_{-0.21}$ & & 1.25/47 & $25.1$\\
    & $1.81^{+0.29}_{-0.25}$ & & $2.00^{+0.24}_{-0.21}$$^b$ & 1.02/47 &$13.9$\\
X19 &$0.61^{+0.47}_{-0.46}$ & $1.75^{+0.33}_{-1.18}$ & &1.02/4&$1.24$\\
X20 & $0.62^{+0.35}_{-0.34}$& & $0.10^{+0.04}_{-0.03}$ & 1.1/6 & $3.18$\\
X21 & $0.32^{+0.05}_{-0.05}$ & $2.53^{+0.16}_{-0.16}$ & & 1.50/82 &$3.88$\\ 
    & $0.08^{+0.02}_{-0.02}$ & & $0.86^{+0.04}_{-0.04}$$^b$ & 2.04/82 &$2.19$\\
    &$0.87^{+0.13}_{-0.25}$& $2.73^{+0.08}_{-0.16}$ & $0.11^{+0.02}_{-0.02}$$^b$&1.25/80&$29.3$\\
    &$0.75^{+0.34}_{-0.18}$ &$1.66^{+0.12}_{-0.22}$& $0.10^{+0.02}_{-0.02}$$^c$&1.24/80&$17.5$\\
X26 & $0.60^{+0.33}_{-0.27}$ &$2.11^{+0.42}_{-0.37}$ && 0.77/14 & $3.49$\\
X30 & $0.51^{+0.33}_{-0.11}$ &$2.85^{+0.97}_{-0.61}$ && 0.97/8 &$0.79$\\
X31 & $0.57^{+0.24}_{-0.27}$ &$3.22^{+1.02}_{-0.64}$ && 0.54/8 &$0.93$\\
X33 &$0.67^{+0.62}_{-0.30}$ &$2.42^{+0.96}_{-0.53}$ && 0.77/7 & $0.97$\\
\hline
\end{tabular}
All quoted uncertainties are at the 90\% confidence level.\\
$^a$ Unabsorbed luminosity ($\times10^{38}$\lum) in 0.5-10 keV,
assuming a distance of 1.8 Mpc.\\
$^b$ Disc blackbody temperature.\\
$^c$ Blackbody temperature from the BMC model.
\end{table}

\subsection{X-ray luminosity function}

We constructed the X-ray point source luminosity function to investigate
the compact object populations of \ic\ (Wu 2001; Kilgard et al. 2002). 
The count rates for all detected
sources were converted into unabsorbed 0.2--12 keV luminosities by assuming
an absorbed power-law model with $N_H=8\times10^{21}$ cm$^{-2}$ and $\alpha=2$.
For the brightest 12 sources (Table 2), we used the luminosities derived
from the best-fitting spectral model.
In addition, we excluded the three foreground stars (X3, X12, X15, X16
and X37). In
Figure 7, we plot the cumulative luminosity function for all sources
(excluding the foreground stars). To estimate the completeness limit,
we used a method described by Kong et al. (2002).
We computed histogram of the number of detected sources against the S/N
to examine the completeness limit; the histogram peaks at S/N $\sim 5$,
corresponding
to $\sim 3\times10^{37}$\lum, and fall off below this. Hence the
luminosity function is complete down to $\sim 3\times10^{37}$\lum.
We fit the
luminosity function with a simple power-law model and determined
the slope via maximum likelihood method on the
differential luminosity function ($dN/dL\propto L^{-\beta}$; e.g., Crawford, Jauncey, \& Murdoch
1970). We note that the exponent for a fit to cumulative function
would be $\Gamma=\beta-1$.
We obtained a slope of $0.55\pm0.07$ for the
cumulative luminosity function, with luminosities
greater than $10^{37}$\lum. There are 12 sources outside the $D_{25}$
disc of \ic\ and they are likely to be contaminated by foreground or
background objects.
Excluding point sources outside the
$D_{25}$ ellipse, the slope ($\Gamma=0.51\pm0.07$) is in good
agreement with the above result. Within the $D_{25}$ ellipse, we
estimated that there
are about 3 background sources accroding to the \chandra\ Deep
Field Surveys (e.g., Brandt et al. 2001; Giacconi et al. 2001).

\begin{figure}
\psfig{file=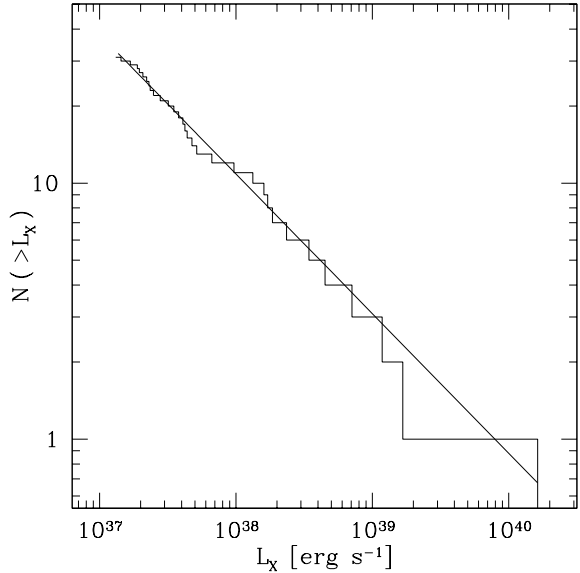,width=8cm}
\caption{Cumulative luminosity function of X-ray point sources in
\ic. The best-fitting model is a power-law with slope $\Gamma=0.55$.}
\end{figure} 

\section{Discussion}

\subsection{Global properties}

Of the 37 detected X-ray sources in the field iof \ic, 25 of them are
inside the $D_{25}$ ellipse. 
Based on the colour-colour diagram and spectral fits, X-ray sources in
\ic\ are likely to be typical X-ray binaries. There are four sources
for which the colours are similar to supernova remnants (see e.g.,
Soria \& Kong 2003), while X20 is
likely to be a luminous supersoft X-ray source (SSS; Di\,Stefano \& Kong
2003). In a detailed study of SSSs in 4 external galaxies, Di\,Stefano
\& Kong (2003) found more than 30 SSSs while only
one SSS in M101 has luminosity $>
10^{39}$\lum (see also Mukai et al. 2003). Recently, Kong \&
Di\,Stefano (2003) also discovered a recurrent luminous SSS in NGC 300
which exceeds $10^{39}$\lum\ during its high state. Therefore, X20 is
one of the few very luminous SSSs found in nearby galaxies. 
Like many SSSs in external
galaxies, it is clear that 
X20 is on the spiral arm of \ic\ (see Figure 3) where young stellar populations
are expected. Comparing the fluxes between the \rosat\ and \xmm\
observations, X20 varies by a factor of 2, suggesting that it is
unlikely to be a supernova remnant. It is therefore possible that X20
is a symbiotic if the accreting object is a white dwarf. However, in
order to achieve such a high luminosity, it is likely that X20
contains a black hole accretor (see Di\,Stefano \& Kong 2003 and Kong
\& Di\,Stefano 2003 for discussion). The spectral
parameters ($kT = 100$ eV, $L_{bol}=1.1\times10^{39}$\lum) is
consistent with a model of an intermediate mass black hole
(Di\,Stefano et al. 2003). The implied black hole mass would be at
least 1200$M_{\odot}$. This is likely to be a lower limit as spectral
hardening, orientation and spin would tend to increase the black hole
mass. An alternative model involves an Compton thick outflow from a
stellar-mass black
hole, accreting near the Eddington limit (King \& Pounds 2003). For
instance, a 100 eV source is consistent with a $10M_\odot$ black hole with
an outflow rate of $10\dot{M}_{Edd}$.

The point source luminosity function of \ic\ can be fit with a
power-law and the slope ($\Gamma=0.55$)
is consistent with other starburst galaxies or spiral disc
($\Gamma\sim 0.5-0.8$; Kilgard et al. 2002; Grimm, Gilfanov, \& Sunyaev 2003). Moreover, 
Grimm, Gilfanov, \& Sunyaev (2002) obtained a slope of
$0.6$ for the high-mass X-ray binaries (HMXBs) in our Galaxy. The similarity
between the luminosity functions of \ic, Galactic HMXBs, and other
starburst galaxies suggests
that the X-ray point source population might be dominated by HMXBs
with continuous star forming activity. More recently, Ranalli,
Comastri \& Setti (2003) determined a linear relation between the star
formation rate (SFR) and total X-ray luminosity (2--10 keV) in various
nearby star-forming galaxies based on \asca\ and \sax\ data. In our \xmm\ observation, the
2--10 keV luminosity of all point sources is $\approx
2.5\times10^{39}$\lum. Applying the relationship derived by Ranalli et
al. (2003),
we obtained a SFR of $\approx 0.5 M_{\odot}$\,yr$^{-1}$. This is 
consistent with the SFR ($0.48 M_{\odot}$\,yr$^{-1}$) obtained from radio
observations (Grimm, et al. 2003). However, for such a
low SFR, the relation between the SFR and X-ray luminosity is
non-linear (Grimm et al. 2003). Following Grimm et al. (2003), the
expected SFR from the observed X-ray luminosity should be between
$0.3-1 M_{\odot}$\,yr$^{-1}$.

\subsection{X-ray emission from the nucleus}

Although it was known that there is X-ray emission from the nucleus of
\ic\ since \einstein, we know very little about its nature. \rosat\
HRI image resolved the nucleus into several X-ray sources, but 
subsequent \asca\ observations do not have the resolving power to
isolate them.  Furthermore, the \rosat\ HRI image indicates that the
nuclear source is extended. 
In the \xmm\ observation, 
X21 is clearly seen and separated from
nearby sources in the \xmm\
image (see Figure 1) and is about $3''$ from the dynamical
galaxy centre. We examined the surface brightness profile of the source
and it is consistent with the instrumental point spread function of a
point source.
Given a relatively large error circle ($\sim 4''$) from
\xmm\ data, it is not clear if X21 is truly associated with the dynamical
centre. Better astrometry by \chandra\ is needed to confirm
the result. From optical and infrared observations, there is a star
cluster in the centre of the galaxy (B$\rm \ddot{o}$ker, van der Marel, \&
Vacca
1999). While a central black hole is not found, an upper
limit of $5\times10^5 M_{\odot}$ can be set. We obtained the X-ray
spectrum of X21. The source is very luminous ($\sim 10^{40}$\lum\ in
0.2--12 keV) and
it requires two-component models to
give a best fit.  The result is similar to many ULXs in external galaxies and
Galactic black hole binaries during the high/soft state. However, the
disc temperature ($kT\approx0.1$ keV) is about 5 to 10 times lower
than expected. Recently, such cool disc systems
have been found in NGC 1313 and
have been interpreted as evidence of intermediate-mass black holes
(Miller et al. 2003). Following Miller et al. (2003), the black hole
mass of X21 is in the range of $220M_{\odot}$ and
$15000M_{\odot}$. If we simply scale the mass according to the
Eddington luminosity, we obtain a minimum black hole mass of $120M_{\odot}$.
However, the powerlaw component of X21 is weak and contributes only $\sim
10\%$ of the total emission. For the two ULXs in NGC 1313, the powerlaw
compontent is 33\% and 63\% of the total X-ray flux (Miller te al. 2003).
The unphysical spectral fit with broken powerlaw model rules out the
possibility of an accreting stellar black hole with beamed relativistic jet
emission (Kaaret et al. 2003). 
If X21 is indeed an intermediate-mass black hole, it
is likely that the source is associated with the star cluster instead
of the galaxy dynamical centre. Ebisuzaki et al. (2001) recently
suggest that intermediate-mass black holes can form in young compact
star clusters through successive mergings of massive stars. This is
suggestive that X21 is an example of this scenario. It is, however, worth
to noting that we cannot rule out a stellar-mass black hole with strong
wind (King \& Pounds 2003; see also \S 4.1).

\subsection{The ultraluminous sources in \ic}

Except for the nucleus, there are two historical ULXs (X7 and X13) in \ic.
We examined the light curves of the individual ULXs and they do
not show significant variability on the timescale of our
observation. The brightest off-nucleus ULX, X7, has a luminosity of
$\sim10^{39}$\lum\ during the \xmm\ observation. Comparing to previous
\asca\ observations, the source is consistent with the low/hard state
(Kubota et al. 2001; see also Kubota, Done \& Makishima
2002). Similarly, the luminosity state of the second brightest ULX, X13, during our \xmm\
observation also corresponds to the low/hard state in the \asca\ observations.
It is worth noting that all previous X-ray observations
derived the luminosity by assuming a distance of 4 Mpc which is
greater than generally acceptable value of 1.8 Mpc (Buta and McCall
1999). We here rescaled all the luminosities assuming 1.8 Mpc. The
smaller implied luminosities of the two brightest ULXs are still
between $\sim 10^{39}-10^{40}$\lum. Hence, a smaller black hole mass
is required. For instance, according to the calculation by Kubota et
al. (2002), X7 would have a black hole mass between $10
M_{\odot}$ and $30 M_{\odot}$.

It is clear from the \xmm\ and \rosat\ (Bregman et al. 1993) images that the
two ULXs with state/intensity changes have several nearby sources. 
The significant poorer spatial resolution of \asca\ (FWHM $> 1'$) and
the large extraction radius ($3'$) used in analysing the \asca\ data (Okada et
al. 1998) imply that the targets suffer confusion problem. 
For example, 
a $3'$ circular region surrounding X7 includes six other sources (X6, X9, X12,
X14, X15, and X16). Comparing the \xmm\ counts of these
sources, about 40\% of the \asca\ counts of X7 in the low state are due to confusion.
Therefore, the luminosities measured by \asca\ are over-estimated
particularly during the low luminosity state.
In fact, the two
ULXs shown in Figure 1 in Kubota et al. (2001) show asymmetric shape
and the centroids have moved during the two observations, indicating
that there are bright nearby X-ray variables.
Comparing the \xmm\ data with the \rosat\ HRI observations, X6
and X9 (not detected in Bregman et al. 1993, but might be marginally
seen in the archival image) varied by a factor of $> 2$. It is
therefore possible that the factor of $\sim 5$ (after correcting
the possible confusion of nearby sources based on \xmm\ data) variability of X7 during
the two \asca\ observations are due to X6 and
X9, and/or other nearby sources if one of these sources varies by at
least a factor of 10. It is more likely that the variability seen
by \asca\ is intrinsic to X7.
In addition, X12 is a foreground star which shows soft spectrum
(see Table 2). If the foreground star has flarings during the
high/soft state in the \asca\ observations, it could contribute
additional soft X-ray emission.  

More recently, X7 was observed by \chandra\ ACIS-S on 2002 April
29 with an exposure time of 10 ks. We therefore use this archival
observation to investigate the contamination of nearby sources in
detail. Within $30''$ of X7, there is no obvious X-ray source and
the nearest point source is about $3.6'$ from X7 \footnote{The
\chandra\ observation was operated in sub-array mode and therefore it
only covered a small part around X7. From our
\xmm\ observation, the nearest point source to X7 is about $50''$
away.}. Any faint source within this area must be fainter than $8\times10^{36}$\lum.
Hence, the contamination of our \xmm\ spectrum should be minimal. We
also examined the \chandra\ spectrum of X7. After correcting the soft
energy degradation and pile-up ($\sim 5\%$), the spectrum can be fit
with a power-law model [$N_H=(5.68\pm0.09)\times10^{21}$cm$^{-2}$,
$\alpha=1.81\pm0.31$] with a reduced $\chi^2$ of 1.0 for 81 degrees of
freedom. The 0.5--10 keV luminosity is $1.6\times10^{39}$\lum, similar
to our \xmm\ observation.

Similarly, for X13, a $3'$ extraction region
covers X11 and X20 as well. In particular, X20 is likely to be a
luminous ($\sim 10^{39}$\lum\ in 0.5--10 keV) 
SSS (see Table 2 and \S\,4.1 for discussion) and
it shows a factor of 2 variability between the \rosat\ and \xmm\
observations. SSSs are well-known to be highly
variables (e.g., Di\,Stefano \& Kong 2003; Kong \& Di\,Stefano 2003)
and therefore the less than
factor of 2 variability of X13 during the two \asca\ observaitons is
easily explained by the existence of a SSS. Note that the SSS has 
relatively high temperature ($kT = 0.1$~keV) and \asca\ should
be able to detect it; the source
luminosity in the \asca\ band (0.7--10 keV) is about $9\times
10^{37}$\lum.
Comparing with the 0.7--2 keV \asca\ count rate of X13 during the high state (Sugiho et
al. 2001),  roughly 25\% of counts are contaminated by the SSS
(assuming the same blackbody model measured by \xmm). Therefore, if the SSS
brightens during the high state by a factor of $> 2$, it could
contribute significant soft photons to result a softer and brighter spectrum. 
It is worth
noting that X13 is asymmetric toward the direction of the
supersoft source, X20, during the high/soft state in 2000 (see Figure
1 in Kubota et al. 2001). We therefore conclude that the
spectral/intensity changes seen in the two \asca\ observations should
be treated with caution. Further monitoring observations with \xmm\
and \chandra\ are required to investigate the state transitions of
ULXs in \ic.

\section{Summary}

We have studied the X-ray point source population in the nearby
starburst galaxy \ic\ with \xmm. Thirty-seven X-ray sources were
detected at a significance of $5\sigma$ or greater. The X-ray point
source luminosity function is in a power-law form and the slope
($\Gamma=0.55$) is consistent with other starburst galaxies and Galactic
HMXBs. Most of the X-ray sources are near the spiral arms, indicating
that the X-ray point sources are dominated by young stellar population
(presumably HMXBs) with continuous star forming activity. For the 12
brightest sources, we extracted the energy spectra and except for the
nucleus, a SSS, and a foreground star, they can be described by a
power-law model. There is one source (X21) about $3''$ from the
dynamical centre, and the spectrum can be fit with a disc blackbody
plus power-law model. The source is very luminous ($\sim 10^{40}$ \lum) and
the disc temperature is low ($kT=0.11$ keV).
It could be an evidence for an intermediate-mass black hole or a
stellar-mass black hole with strong wind. We also
investigate the intensity/spectral changes of the two famous ULXs but
our \xmm\ observation suggests that the changes could be simply due to
confusion of \asca\ observations.

\section*{Acknowledgments}
I thank Rosanne Di\,Stefano and Phil Kaaret for useful discussion.
This work was supported by NASA under an LTSA grant, NAG5-10705, and by
the Croucher Foundation. This work is based on observations obtained with
\xmm, an ESA mission with
instruments and contributions directly funded by ESA member states and
the US (NASA).

\label{lastpage}
\end{document}